# Unexpected Giant Microwave Conductivity in a Nominally Silent BiFeO$_3$ Domain Wall

*Yen-Lin Huang, Lu Zheng, Peng Chen, Xiaoxing Cheng, Shang-Lin Hsu, Tiannan Yang, Xiaoyu Wu, Louis Ponet, Ramamoorthy Ramesh, Long-Qing Chen, Sergey Artyukhin, Ying-Hao Chu, Keji Lai*[*]*

Dr. Y.-L. Huang, L. Zheng, Dr. X. Wu, Dr. K. Lai

Department of Physics, University of Texas at Austin, Austin, TX 78712, USA

Dr. Y.-L. Huang, Dr. Y.-H. Chu

Department of Materials Science and Engineering, National Chiao Tung University, Hsinchu, Taiwan

Dr. P. Chen, L. Ponet, Dr. S. Artyukhin

Quantum Materials Theory, Istituto Italiano di Tecnologia, Genova, Italy

X. Cheng, T. Yang, Dr. L.-Q. Chen

Department of Materials Science and Engineering, The Pennsylvania State University, University Park, PA 16082, USA

Dr. Y.-L. Huang, Dr. R. Ramesh

Department of Materials Science and Engineering & Department of Physics, University of California, Berkeley, CA 94720, USA

Dr. Shang-Lin Hsu

Materials Sciences Division, Lawrence Berkeley Laboratory, Berkeley CA 94720, USA

**Abstract**

Nanoelectronic devices based on ferroelectric domain walls (DWs), such as memories, transistors, and rectifiers, have been demonstrated in recent years. Practical high-speed electronics, on the other hand, usually demand operation frequencies in the giga-Hertz (GHz) regime, where the effect of dipolar oscillation is important. In this work, an unexpected giant GHz conductivity on the order of $10^3$ S/m is observed in certain $BiFeO_3$ DWs, which is about 100,000 times greater than the carrier-induced dc conductivity of the same walls. Surprisingly, the nominal configuration of the DWs precludes the ac conduction under an excitation electric field perpendicular to the surface. Theoretical analysis shows that the inclined DWs are stressed asymmetrically near the film surface, whereas the vertical walls in a control sample are not. The resultant imbalanced polarization profile can then couple to the out-of-plane microwave fields and induce power dissipation, which is confirmed by the phase-field modeling. Since the contributions from mobile-carrier conduction and bound-charge oscillation to the ac conductivity are equivalent in a microwave circuit, the research on local structural dynamics may open a new avenue to implement DW nano-devices for RF applications.

When a multi-domain ferroelectric crystal is placed in an ac electric ($E$) field, the electrostatic coupling between the spontaneous polarization and the external field may lead to periodic oscillation of the domain walls (DWs).[1-3] Because of the inertia of dipole moments and frictional effects in the material, such DW motion gives rise to collective dielectric loss at the microwave frequency ($f$), which limits the use of bulk ferroelectrics for RF tunable devices.[4] The situation could be different, however, if we can fully understand the dielectric response of individual DWs and address them separately in electronic devices. For instance, one may take advantage of the ac conductivity of DWs due to the dipolar loss $\sigma_1^{ac} = \omega\varepsilon''$, where $\omega = 2\pi f$ and $\varepsilon''$ is the imaginary part of the permittivity, to function as device interconnects or high-$f$ waveguides. It should be noted that in the past decade, DW conduction down to zero-frequency due to mobile carriers has been discovered in many ferroelectrics.[5,6] At microwave frequencies, the total ac conductivity $\sigma^{ac}$ is the sum of $\sigma_1^{ac}$ and $\sigma_2^{ac}$, where the Drude-like free-carrier contribution $\sigma_2^{ac} = \sigma^{dc}$ is essentially the same at both dc and GHz.[4] On the other hand, while mobile-carrier conduction and bound-charge oscillation both contribute to the ac conduction, they are fundamentally different physical processes and should be analyzed separately. It is therefore imperative to obtain a unified picture on these two mechanisms in ferroelectrics DWs, which is crucial for their applications in nanoelectronics. [7-11]

The giga-Hertz (GHz) response of ferroelectric DWs has been spatially resolved by microwave impedance microscopy (MIM).[12,13] For weakly charged walls in KNbO$_3$ crystals[14] and transiently formed charged walls in lead zirconate (PZT) thin films,[15] it was shown that the DW ac conductivity is dominated by the charge carriers, i.e., $\sigma^{ac} \approx \sigma^{dc} = \sigma_2^{ac}$. In contrast, the MIM result on hexagonal manganites (h-$R$MnO$_3$)[16] and ferrites (h-$R$FeO$_3$)[17] depends strongly on the DW orientation. On the (001) surface with uncharged 180° DWs, $\sigma_{DW}^{ac}$ is 4 ~ 6 orders of magnitude higher than $\sigma_{DW}^{dc}$, suggesting that the ac *E*-field from the MIM tip can effectively drive the DW oscillation.[16] For charged walls on the (110) surface, however, the MIM *E*-field does not favor the polarization on either side of the wall. As a result, no enhancement of $\sigma_{DW}^{ac}$ over the dc value was observed.[16] In this work, we report the nanoscale microwave conductivity imaging on BiFeO$_3$ domain walls that nominally do not vibrate under out-of-plane ac *E*-fields. Surprisingly, while $\sigma^{ac} \approx \sigma^{dc} = \sigma_2^{ac}$ is indeed observed in the control sample with vertical 71° walls, the effective GHz conductivity of the inclined 71° walls, which nominally should not oscillate under the out-of-plane excitation electric field, is $10^5$ times higher than its $\sigma^{dc}$. Using a simplified Ginzburg-Landau theory,[18] we find that the inclination of the wall leads to an asymmetric profile of the out-of-plane polarization, which is responsible for its vibration under the ac *E*-field. The dielectric loss due to displacement current at the tilted DWs is also confirmed by the phase-field modeling. Our results highlight the importance of local symmetry in the structural dynamics of ferroelectric DWs. In analogy to the proposed application of DW dc conduction in memories and transistors [7-11], the giant ac conductivity of certain ferroelectric DWs may also be utilized for radio-frequency nanoelectronics such as microwave interconnects or waveguides.

The main sample in this work – 150 nm BiFeO$_3$ (BFO) on 3 nm conductive SrRuO$_3$ (SRO) thin film, hereafter referred to as Sample A – was epitaxially grown on DyScO$_3$ (DSO) substrates by pulse laser deposition.[19,20] BFO is one of the most promising multiferroic materials[21] and its DW properties are technologically important. The coherent growth is facilitated by the close match between the DSO lattice on the (110)$_O$ orthorhombic surface and the BFO lattice on the (001)$_C$ pseudocubic surface. As illustrated in **Figure 1a**, the BFO film displays an array of stripe domains oriented along the [010]$_C$ direction.[19,20] The 71° DWs, categorized by the angle between the polarization vector of two neighboring domains (inset of Figure 1a), are uncharged since the polar discontinuity is parallel to the wall plane. Moreover, the change of polarization vector across the wall lies in the plane of film surface. Consequently, an out-of-plane oscillating *E*-field, which is most relevant for thin-film devices with bottom electrodes, should not couple to the DWs here.

In order to study the GHz dielectric response of Sample A, we performed the MIM experiment,[12,13] where the microwave signal is delivered to the center conductor of a shielded cantilever probe.[22] By amplifying and demodulating the reflected signal, the imaginary (MIM-Im) and real (MIM-Re) parts of the tip-sample admittance can be spatially resolved. Before the MIM imaging, the probe and electronics are calibrated by measuring by measuring standard samples, e.g., Al dots on sapphire substrates.[13] Figure 1b shows the atomic-force microscopy (AFM), in-plane piezo-force microscopy (PFM), and MIM (*f* = 1 GHz) images of Sample A. With virtually no crosstalk to the surface topography, the stripe domains are clearly visualized in the PFM image. The microwave images, on the other hand, exhibit strong signals at the walls. Similar behaviors were also observed at other frequencies ranging from 20 MHz to 5 GHz (Supporting Information S1). The frequency dependence is comparable to that observed in h-*R*MnO$_3$ [16], indicative of similar physical origins between the two systems. In the following, we will focus on the 1 GHz data with good signal-to-noise ratios in both MIM channels.

The MIM contrast between DWs and domains in Sample A can be vividly seen from the line profiles in Figure 1c. To quantify the result, we carried out finite-element analysis (FEA, Supporting Information S2) to compute the complex tip-sample admittance and convert them to the MIM signals.[12] Figure 1d shows the simulated results as a function of the effective DW ac conductivity. Since the measured DW signals in the MIM-Im channel is ~ 10 times stronger than that in MIM-Re (Figure 1c), a comparison to the numerical simulation with the ratio MIM-Re/MIM-Im of ~ 0.1 [16] indicates that $\sigma_{DW,A}^{ac}$ is ~ 10$^3$ S/m, the highest value reported in BFO to date.[5,23] In contrast, the conductive AFM (c-AFM) result shows that the carrier-induced dc conductivity $\sigma_{DW,A}^{dc}$ of these charge-neutral walls is only $10^{-2}$ S/m (Supporting Information S3). It should be noted that while defects may accumulate near ferroelectric DWs, hopping conduction of weakly localized states at finite frequencies [24-26] cannot explain the ~ $10^5$ times higher $\sigma_{DW,A}^{ac}$ over $\sigma_{DW,A}^{dc}$. Such a drastic difference implies that the microwave response of the inclined 71° DWs is dominated by the collective dipolar loss rather than the Ohmic loss. As discussed before, the vibrational motion of this DW, if exists at all, is in principle decoupled from the out-of-plane microwave fields (inset of Figure 1d, see Supporting Information S2). The excitation of such a nominally silent mode in our MIM experiment is thus unexpected and highly nontrivial.

Before further investigating the DW vibration in Sample A, it is instructive to evaluate the result of a control sample, 50 nm BFO / 30 nm SRO thin film on $(110)_C$ cubic $SrTiO_3$ substrate, hereafter referred to as Sample B. As depicted in **Figure 2a**, the $(110)_C$-oriented BFO film exhibits an irregular domain pattern that is consistent with our earlier work (Supporting Information S3).[23] DWs in this sample are also categorized as 71° walls. The two polarization vectors in neighboring domains span from the 'head-to-head' (HtH), neutral, to the 'tail-to-tail' (TtT) configurations. The maximum DW dc conductivity due to free-carrier conduction is ~ 3 S/m in the HtH configuration (Supporting Information S3), consistent with previous reports.[27,28] In Figure 2b, the contours of ferroelectric domains obtained from the in-plane PFM are overlaid on the MIM images taken at $f$ = 1 GHz. Neither domain nor DW contrast appears in the MIM-Im channel, whereas broken sections of DWs are seen in MIM-Re. Detailed analysis (Supporting Information S4) shows that these isolated segments coincide with the HtH sections of DWs. Our transmission electron microscopy (TEM) (Supporting Information S5) and phase-field simulation (Supporting Information S6) results indicate that the HtH sections of DWs in Sample B are perpendicular to the film surface, as illustrated in the inset of Figure 2c. Based on the FEA simulation in Figure 2c, $\sigma_{DW}^{dc}$ and $\sigma_{DW}^{ac}$ are essentially the same for the HtH walls in Sample B. In other words, neither hopping conduction [24–26] nor DW vibration is significant and the microwave response of vertical 71° DWs can be fully accounted for by the mobile-carrier contribution.

The major difference between the two BFO samples is the orientation of DWs with respect to the surface. In an infinite bulk sample with one DW, the diagonal mirror plane indicated by the dashed line in **Figure 3a** is a symmetry operation. As a result, the nominal orientation of DWs in Sample A is tilted 45° away from the surface normal. In thin films with a finite thickness, however, the strain near the surface can be relaxed and violation of the compatibility constraint [29] does not lead to infinite elastic energy. The stress near the surface is therefore imbalanced on two sides of the wall. Using a simplified Ginzburg-Landau model,[18] one can show that the DW bends towards the normal direction near the surface.[30] We note that the bending of 71° DWs near the surface of BFO thin films has been experimentally observed by TEM and PFM studies.[31,32] Figure 3a shows the spatial distribution of $P_z$ (out-of-plane component of the polarization) in the cross-section of the film, as simulated by electromechanical finite-element method.[30] Because of the large spontaneous polarization in BFO,[21] the imbalance of $P_z$ across the wall is comparable to polarization change across the 180° walls in h-$R$MnO$_3$[16] and h-$R$FeO$_3$.[17] The asymmetric $P_z$ profile can now couple to the out-of-plane MIM $E$-fields and induce DW vibration in Sample A. In contrast, since DWs in Sample B are perpendicular to the film surface, no asymmetry in the stress/strain or polarization is induced around the wall (Supporting Information S7), thus the absence of DW dielectric loss.

The excitation of DW vibration in Sample A can be validated by our dynamical phase-field model,[33] where the time-dependent response of the polarization vector is computed using the polarization dynamics equation [34,35] (Supporting Information S8). Figure 3b shows a snapshot of $\partial P_z/\partial t$ at $t = T/4$ when a sinusoidal $E$-field ($\propto \sin 2\pi t/T$) is uniformly applied between the top surface and the conductive substrate. The bipolar line shape around the wall clearly manifests the DW sliding [16] under the out-of-plane ac $E$-field. To understand the influence of DW vibration on electrical energy loss, we performed a series of simulations by moving a tip-induced potential profile across the sample surface. Figure 3c shows the spatial distribution of power density $\partial \boldsymbol{P}/\partial t \cdot \boldsymbol{E}$ integrated over one period when the tip is on top of DW and away from the wall. The effective ac conductivity, estimated from the spatial summation of the time-averaged value $(\partial \boldsymbol{P}/\partial t \cdot \boldsymbol{E})/E^2$, is plotted as a function of the tip location in Figure 3d. The appearance of strong $\sigma_1^{ac}$ at the DW demonstrates that our simple 2D dynamical phase-field simulations capture the essential physics behind the experimental observation in Sample A.

In summary, we discover the excitation of a nominally silent mode in $BiFeO_3$ domain walls by the out-of-plane electric field from a microwave probe. The effective ac conductivity of such inclined 71° DWs is about $10^5$ times greater than the dc value, signifying the predominance of bound-charge oscillation over mobile-carrier conduction in the sample. Our analysis based on the electrostriction effect shows that the out-of-plane polarization is imbalanced around the wall. Phase-field simulation further indicates that such an asymmetric polarization profile can couple to the ac electric field from the tip, resulting in strong power dissipation at the DW. We emphasize that, while the physical origin is different, the ac conductivity due to dipolar loss is equivalent to that from the electron conduction for microwave electronics. The collective vibration under an out-of-plane electric field is localized perpendicular to the wall but free to propagate along the DW plane,[16,30] which may be utilized as, e.g., high-$f$ waveguides. In that sense, this work represents an important step towards implementing DW nanoelectronics for RF applications.

**Experimental Section**

*Material synthesis*: Both BFO thin-film samples in this work were grown by the pulsed laser deposition (PLD) technique. Sample A (150 nm BiFeO$_3$ on 3 nm SrRuO$_3$) was deposited on single-crystal (110)$_O$ DyScO$_3$ substrate and Sample B (50 nm BiFeO$_3$ on 30 nm SrRuO$_3$) on (110)$_C$ SrTiO$_3$, respectively. The growth of BiFeO$_3$ and SrRuO$_3$ layers was performed at 700°C under an oxygen pressure of 100 mTorr with a laser fluence of ~1 J/cm$^2$ and a repetition rate of 5 Hz. After the growth, the films were cooled to room temperature in 500 Torr of oxygen at a rate of 5 °C/min to optimize the oxidation level.

*Imaging experiments and finite-element analysis*: The conductive atomic-force microscopy (C-AFM), piezo-force microscopy (PFM), and microwave impedance microscopy (MIM) experiments were carried out on a commercial AFM platform (XE-70, ParkAFM). The electrically shielded microwave cantilever probes[22] are commercially available from PrimeNano Inc. The two output channels of MIM correspond to the real and imaginary parts of the local sample admittance, from which the effective ac conductivity of the sample can be deduced. Numerical simulation of the MIM signals was performed by the finite-element analysis (FEA) [16] software COMSOL4.4. Details of the FEA on the two samples are included in Supporting Information S2.

*Dynamic phase-field simulation*: A phase-field model taking into account the polarization dynamics was used to simulate the domain and domain wall response in the BFO thin films under an ac electric field [14,34,35]. The 2D simulation we performed has 512 grids in the *x*-direction and 64 grids in the *z*-direction, with each grid representing 0.4 nm. Along the *z*-direction, there are 10 grids for the conductive substrate, 50 grids for the film, and 4 grids as vacuum. The DW of interest is roughly at the center of the simulated geometry. Details of the phase-field modeling are included in Supporting Information S8.

**Supporting Information**

Supporting Information is available from the Wiley Online Library or from the author.

**Acknowledgements:** Y.-L.H., L.Z., and K.L. were supported by NSF Award No. DMR-1707372 and the Welch Foundation Grant F-1814. T.Y. was supported by NSF Award No. DMR-1744213. X.C and L.-Q. C. was supported by the U.S. Department of Energy, Office of Basic Energy Sciences, Division of Materials Sciences and Engineering under Award DE-FG02-07ER46417. S.-L.H. was supported by the U.S. Department of Energy, Office of Science, Office of Basic Energy Sciences, Materials Sciences and Engineering Division, under contract number DE-AC02-05-CH11231 (Quantum Materials program KC2202).

**Author contributions:** Y.-L.H., L.Z., P.C., and X.C. contributed equally to this work. Y.-H.C. and K.L. conceived and designed the experiments. Y.-L.H. grew the materials. Y.-L.H. and L.Z. performed the MIM experiment and numerical analysis. P.C., L.P. and S.A. performed the theoretical studies. X.C. and T.Y. performed the phase-field modeling. S.-L.H. performed the TEM imaging. All authors were involved in the discussion of results and edited the manuscript.

**Conflict of interest:** K.L. holds a patent on the MIM technology, which is licensed to PrimeNano Inc. for commercial instrument. The terms of this arrangement have been reviewed and approved by the University of Texas at Austin in accordance with its policy on objectivity in research. The remaining authors declare no competing financial interests.

# References

[1]    C. Kittel, *Phys. Rev.* **1951**, 83, 458.

[2]    M. Maglione, R. Böhmer, A. Loidl, U. T. Höchli, *Phys. Rev. B* **1989**, 40, 11441.

[3]    G. Arlt, U. Böttger, S. Witte, *Ann. Phys.* **1994**, 3, 578.

[4]    S. Gevorgian, *Ferroelectrics in Microwave Devices, Circuits and Systems: Physics, Modeling, Fabrication and Measurements* (Springer, 2009).

[5]    J. Seidel, L. W. Martin, Q. He, Q. Zhan, Y. H. Chu, A. Rother, M. E. Hawkridge, P. Maksymovych, P. Yu, M. Gajek, N. Balke, S. V. Kalinin, S. Gemming, F. Wang, G. Catalan, J. F. Scott, N. A. Spaldin, J. Orenstein, and R. Ramesh, *Nat. Mater.* **2009**, 8, 229.

[6]    R. K. Vasudevan, W. Wu, J. R. Guest, A. P. Baddorf, A. N. Morozovska, E. A. Eliseev, N. Balke, V. Nagarajan, P. Maksymovych, and S. V. Kalinin, *Adv. Funct. Mater.* 2013, 23, 2592.

[7]    G. Catalan, J. Seidel, R. Ramesh, and J. F. Scott, *Rev. Mod. Phys.* **2012**, 84, 119.

[8]    J. A. Mundy, J. Schaab, Y. Kumagai, A. Cano, M. Stengel, I. P. Krug, D. M. Gottlob, H. Doganay, M. E. Holtz, R. Held, Z. Yan, E. Bourret, C. M. Schneider, D. G. Schlom, D. A. Muller, R. Ramesh, N. A. Spaldin, and D. Meier, *Nat. Mater.* **2017**, 16, 622.

[9]    P. Sharma, Q. Zhang, D. Sando, C. H. Lei, Y. Liu, J. Li, V. Nagarajan, and J. Seidel, *Sci. Adv.* **2017**, 3, e1700512.

[10]    J. Jiang, Z. L. Bai, Z. H. Chen, L. He, D.W. Zhang, Q. H. Zhang, J. A. Shi, M. H. Park, J. F. Scott, C. S. Hwang, and A. Q. Jiang, *Nat. Mater.* **2017**, 17, 49.

[11]    J. Schaab, S. H. Skjærvø, S. Krohns, X. Dai, M. E. Holtz, A. Cano, M. Lilienblum, Z. Yan, E. Bourret, D. A. Muller, M. Fiebig, S. M. Selbach, and D. Meier, *Nat. Nanotech.* **2018**, 13, 1028.

[12]    K. Lai, W. Kundhikanjana, M. Kelly, and Z. X. Shen, *Rev. Sci. Instrum.* **2008**, 79, 063703.

[13]    K. Lai, W. Kundhikanjana, M. Kelly, and Z. X. Shen, *Appl. Nanosci.* **2011**, 1, 13.

[14]    T. T. A. Lummen, J. Leung, A. Kumar, X.Wu, Y. Ren, B. K. Van Leeuwen, R. C. Haislmaier, M. Holt, K. Lai, S. V. Kalinin, and V. Gopalan, *Adv. Mater.* **2017**, 29, 1700530.

[15]    A. Tselev, P.Yu, Y. Cao, L. R. Dedon, L.W. Martin, S.V. Kalinin, and P. Maksymovych, *Nat. Commun.* **2016**, 7, 11630.

[16] X. Wu, U. Petralanda, L. Zheng, Y. Ren, R. Hu, S.-W. Cheong, S. Artyukhin, and K. Lai, *Sci. Adv.* **2017**, 3, e1602371.

[17] X. Wu, K. Du, L. Zheng, D. Wu, S.-W. Cheong, and K. Lai, *Phys. Rev. B* **2018**, 98, 081409(R).

[18] S. Nambu and D. A. Sagala, *Phys. Rev. B* **1994,** 50, 5838.

[19] Y.-H. Chu, Q. Zhan, L. W. Martin, M. P. Cruz, P.-L. Yang, G. W. Pabst, F. Zavaliche, S.-Y. Yang, J.-X. Zhang, L.-Q. Chen, D. G. Schlom, I.-N. Lin, T.-B. Wu, and R. Ramesh, *Adv. Mater*. **2006**, 18, 2307.

[20] Y.-Hao Chu, Q. He, C.-H. Yang, P. Yu, L. W. Martin, P. Shafer, and R. Ramesh, *Nano Lett*. **2009**, 9, 1726.

[21] J. Wang, J. B. Neaton, H. Zheng, V. Nagarajan, S. B. Ogale, B. Liu, D. Viehland, V. Vaithyanathan, D. G. Schlom, U. V. Waghmare, N. A. Spaldin, K. M. Rabe, M. Wuttig, and R. Ramesh, *Science* **2003**, 299, 1719.

[22] Y. Yang, K. Lai, Q. Tang, W. Kundhikanjana, M. A. Kelly, K. Zhang, Z.-x. Shen, and X. Li, *J. Micromech. Microeng*. **2012**, 22, 115040.

[23] Y.-H. Chu, M. P. Cruz, C.-H. Yang, L. W. Martin, P.-L. Yang, J.-X. Zhang, K. Lee, P. Yu, L.-Q. Chen, and R. Ramesh, *Adv. Mater*. **2007**, 19, 2662.

[24] M. Pollak, and T. H. Geballe, *Phys. Rev*. **1961**, 122, 1742.

[25] P. Lunkenheimer, V. Bobnar, A. V. Pronin, A. I. Ritus, A. A. Volkov, and A. Loidl, *Phys. Rev. B* **2002**, 66, 052105.

[26] P. Lunkenheimer, and A. Loidl, *Phys. Rev. Lett*. **2003**, 91, 207601.

[27] A. Crassous, T. Sluka, A. K. Tagantsev, and N. Setter, *Nat. Nanotech*. **2015**, 10, 614.

[28] F. Bai, G. Yu, Y. Wang, L. Jin, H. Zeng, X. Tang, Z. Zhong, and H. Zhang, *Appl. Phys. Lett*. **2012**, 101, 092401.

[29] J. Fousek, and V. Janovec, *J. Appl. Phys*. **1969**, 40, 135.

[30] P. Chen, L. Ponet, K. Lai, and S. Artyukhin, arXiv:1907.12989.

[31]    L. Li, J. R. Jokisaari, Y. Zhang, X. Cheng, X. Yan, C. Heikes, Q. Lin, C. Gadre, D. G. Schlom, L.-Q. Chen, and X. Pan, *Adv. Mater*. **2018**, 30, 1802737.

[32]    Y. Zhang, H. Lu, X. Yan, X. Cheng, L. Xie, T. Aoki, L. Li, C. Heikes, S. P. Lau, D. G. Schlom, L. Chen, A. Gruverman, and X. Pan, *Adv. Mater*. **2019**, 31, 1902099.

[33]    H. Akamatsu, Y. Yuan, V. A. Stoica, G. Stone, T. Yang, Z. Hong, S. Lei, Y. Zhu, R. C. Haislmaier, J. W. Freeland, L.-Q. Chen, H. Wen, and V. Gopalan, *Phys. Rev. Lett*. **2018**, 120, 096101.

[34]    Y. L. Li, S. Y. Hu, Z. K. Liu, and L. Q. Chen, *Appl. Phys. Lett*. **2001**, 78, 3878.

[35]    Y. L. Li, S. Y. Hu, Z. K. Liu, and L. Q. Chen, *Appl. Phys. Lett*. **2002**, 81, 427.

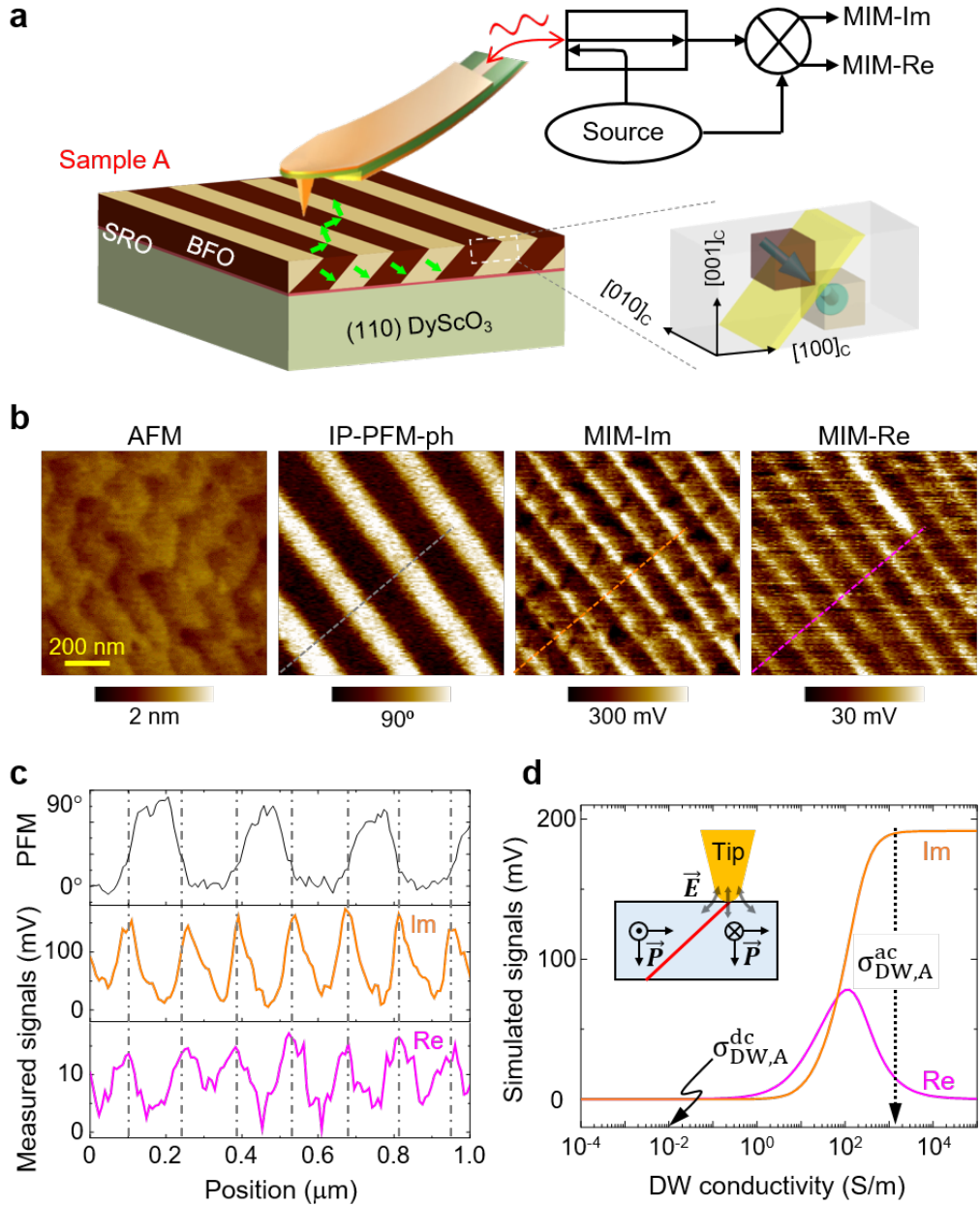

**Figure 1.** Microwave imaging on inclined BiFeO$_3$ domain walls. (**a**) Schematics of the MIM setup and domain structures in Sample A. The microwave signal at 1 GHz is sent to the center conductor of a shielded cantilever probe and the reflected signal is demodulated to form the MIM-Im/Re images. The polarization vectors (green arrows) of the BFO domains are labeled on both the surface and cross-section of the film. The inset shows a close-up view of the polarization direction on both sides of the DW (yellow slab). (**b**) From left to right: AFM, phase image of in-plane PFM, and MIM-Im/Re images on Sample A. All images (1 μm × 1 μm) were taken at the same location.

(**c**) PFM and MIM line profiles across the dash lines in (b). The dash-dotted lines indicate that the MIM signals indeed occur at the domain boundaries. Note that the DW signals are ~ 10 times greater in the MIM-Im channel than that in MIM-Re. (**d**) Simulated MIM signals as a function of the effective DW ac conductivity in Sample A. The corresponding $\sigma_{DW}^{ac}$ that matches the measured signals is marked by the dashed arrow. The DW dc conductivity is also indicated for comparison. The inset illustrates the tip-sample geometry, as well as the directions of the MIM *E*-fields and polarization vectors.

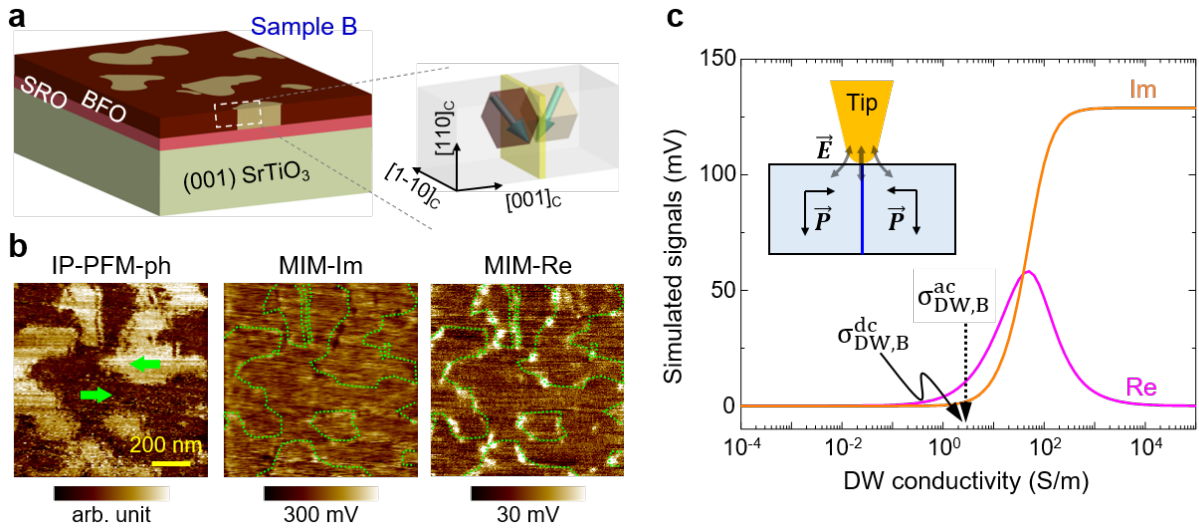

**Figure 2.** MIM result on Sample B. (**a**) Schematics of the layer and domain structures of Sample B. The inset shows the polarization vectors on both sides of the DW (yellow slab) in the 'head-to-head' (HtH) configuration. (**b**) From left to right: phase image of in-plane PFM and MIM-Im/Re images on Sample B. Green dotted lines in the MIM images mark the contour of domains determined from the PFM data. (**c**) Simulated MIM signals as a function of the effective DW ac conductivity in Sample B. Solid and dashed arrows indicated the dc and ac conductivity of the HtH DWs in this sample, respectively. The inset illustrates the tip-sample geometry, as well as the directions of the MIM $E$-fields and polarization vectors.

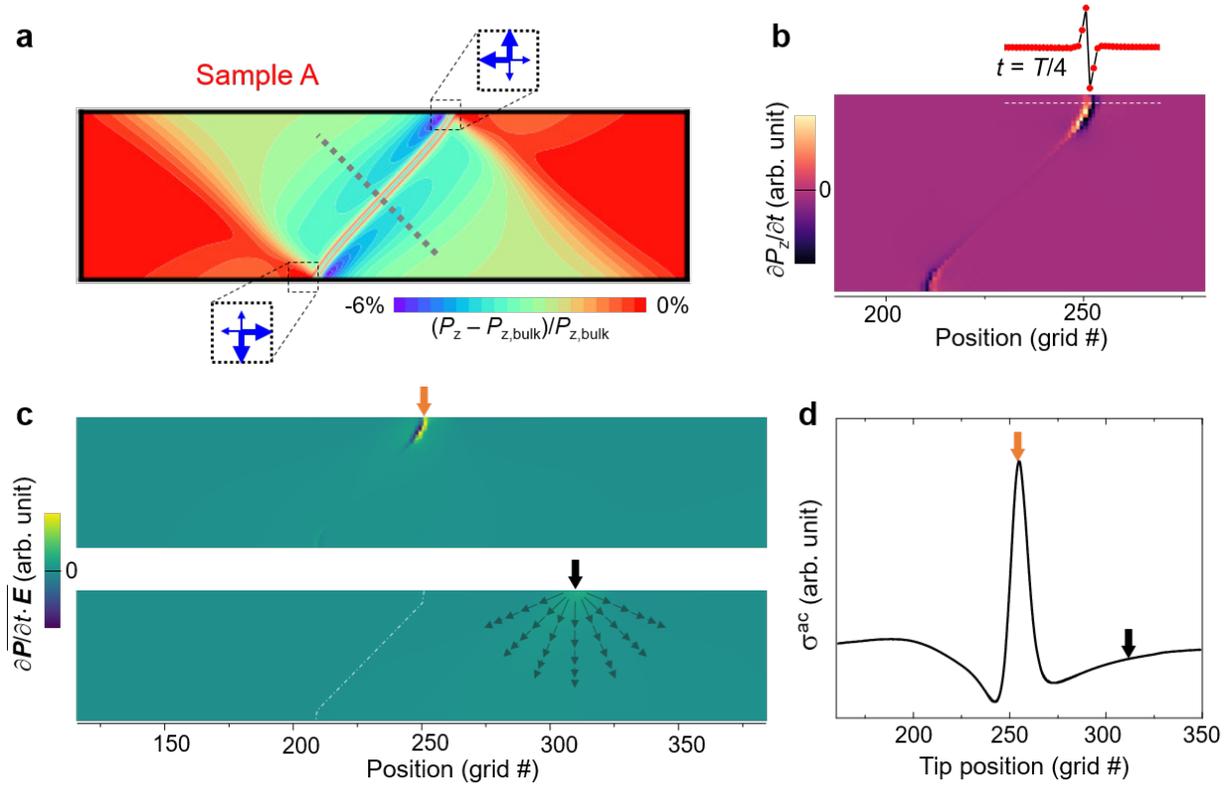

**Figure 3. Theoretical analysis of DW oscillation in Sample A**. (**a**) Simulated out-of-plane polarization $P_z$ in Sample A, showing the imbalanced $P_z$ on two sides of the wall. Note that the tilt angle in the middle of the wall is slightly different from 45° because of the finite film thickness in the simulation. Arrows in the dotted boxes represent the strength of stress along different directions. The net effect causes the DW to bend towards the surface normal. (**b**) Simulated $\partial P_z/\partial t$ at $t = T/4$ when a uniform ac $E$-field is applied between the top surface and the substrate. The DW sliding under oscillating electric fields is evident from the bipolar line profile across the wall. (**c**) Spatial distribution of the time-averaged dielectric loss density when a tip-like potential is placed (top) on top of and (bottom) away from the DW. The tip position is indicated by the orange and black arrows. The DW is depicted by the white dash-dotted line in the bottom image. Gray arrows represent the $E$-field from the tip. (**d**) Simulated ac conductivity as a function of the tip position, showing a sharp peak when the tip scans across the DW.

# Supporting Information

**Unexpected Giant Microwave Conductivity in a Nominally Silent BiFeO$_3$ Domain Wall**

*Yen-Lin Huang, Lu Zheng, Peng Chen, Xiaoxing Cheng, Shang-Lin Hsu, Tiannan Yang, Xiaoyu Wu, Louis Ponet, Ramamoorthy Ramesh, Long-Qing Chen, Sergey Artyukhin, Ying-Hao Chu, Keji Lai[*]*

## S1. Frequency dependence of DW signals in Sample A.

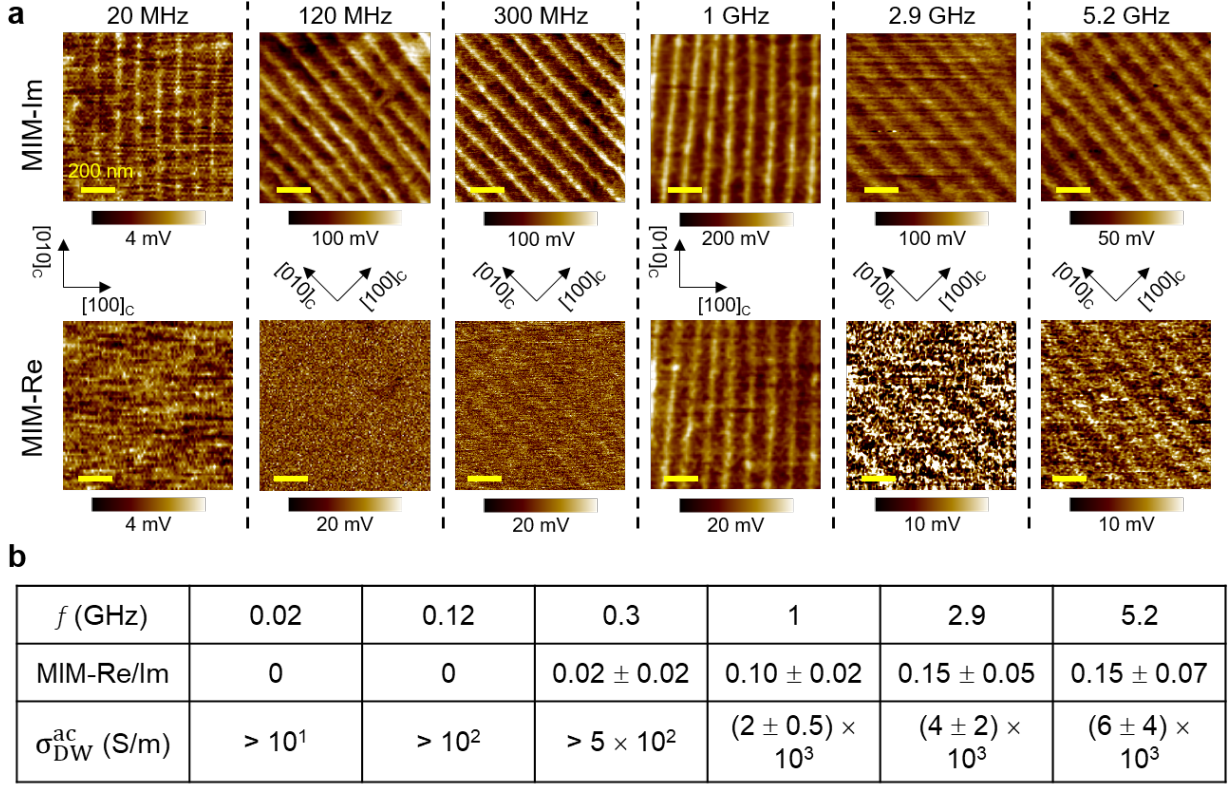

**Figure S1**. (**a**) MIM-Im and MIM-Re images taken at different excitation frequencies. All scale bars are 200 nm. (**b**) DW ac conductivity at various frequencies estimated from the ratio of MIM-Re/Im contrast. The error bars of $\sigma_{DW}^{ac}$ is estimated from the uncertainty of MIM-Re/Im.

The frequency-dependent MIM images of the inclined DWs in Sample A are shown in Figure S1a. Note that the MIM signals are symmetric around the DWs. Since the data were taken using different sets of MIM electronics, it is not possible to compare the absolute signal strengths at different frequencies. It is clear that the DWs exhibit strong MIM-Im signals from 20 MHz to 5 GHz. Following the procedure in Ref. [16] in the main text, we are calculating the ratio of MIM-Re/MIM-Im contrast, which is much less sensitive to the tip diameter and tip-sample contact area in the actual experiment, to perform quantitative analysis. Because of the very weak MIM-Re signals, it is difficult to determine the effective DW ac conductivity in a quantitative manner using the response curves in Figure 1d in the main text. In the table of Figure S1b, we provide the simulated $\sigma_{DW}^{ac}$ values based on the FEA curve. Due to the poor signal-to-noise ratio in the MIM-Re data, the results only serve as order-of-magnitude estimates. The overall characteristics are similar to that observed in h-$R$MnO$_3$ [S1].

## S2. Finite-element analysis of the MIM data.

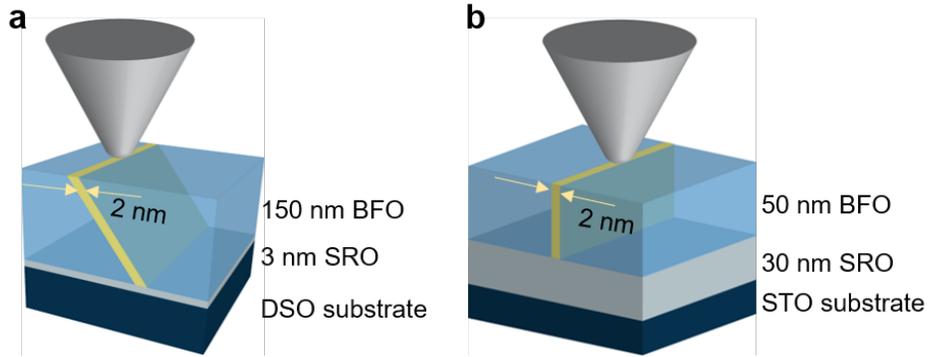

**Figure S2.** Tip-sample geometry for the FEA modeling in (**a**) Sample A and (**b**) Sample B.

Quantitative analysis of the MIM data was carried out by finite-element modeling [S2] using COMSOL 4.4. The program can directly compute the admittance between the tip and the ground based on the tip-sample geometry (Figure S2). Due to the lack of axisymmetry when a DW is involved, 3D modeling is needed for this work. Dimensions of the sample structure are labeled in Figure S2. The DWs are modeled as a thin slab with a thickness of 2 nm. Parameters for the FEA are as follows: tip radius $r = 50$ nm; tip height $h = 500$ nm; half-cone angle of the pyramidal tip $\theta = 22°$; tip-sample distance (in order to avoid a divergent signal when $\sigma$ is very large) $t = 1$ nm; dielectric constant of bulk BFO $\varepsilon_r = 30$. Note that for Sample A, DWs near the surface bend towards the normal direction (see main text). However, the effect on the FEA result is small and not considered here.

From the tip-sample configuration in Figure S2, it is also obvious that the component of tip *E*-field perpendicular to the sample surface is responsible for the MIM signals. First of all, both BiFeO$_3$ thin films (150 nm for Sample A and 50 nm for Sample B) have a metallic bottom layer of SrRuO$_3$. Since the MIM mostly probes the region underneath the tip (diameter ~ 100 nm), the configuration is similar to a parallel-plate capacitor with the tip as the top electrode, where the *E*-field is largely perpendicular to the film. Secondly, the parallel component of the *E*-field is heavily dependent on the exact tip shape at its apex. If that is the dominant effect, the MIM signal would be unrepeatable from tip to tip, which is not observed in our experiment. Finally, from the simultaneously acquired AFM images, the surface roughness of both samples is within 1 ~ 2 nm, much smaller than the film thickness or the tip size. The effect of surface topography on the *E*-field direction is therefore minimal.

## S3. C-AFM/PFM data of the two BFO thin films.

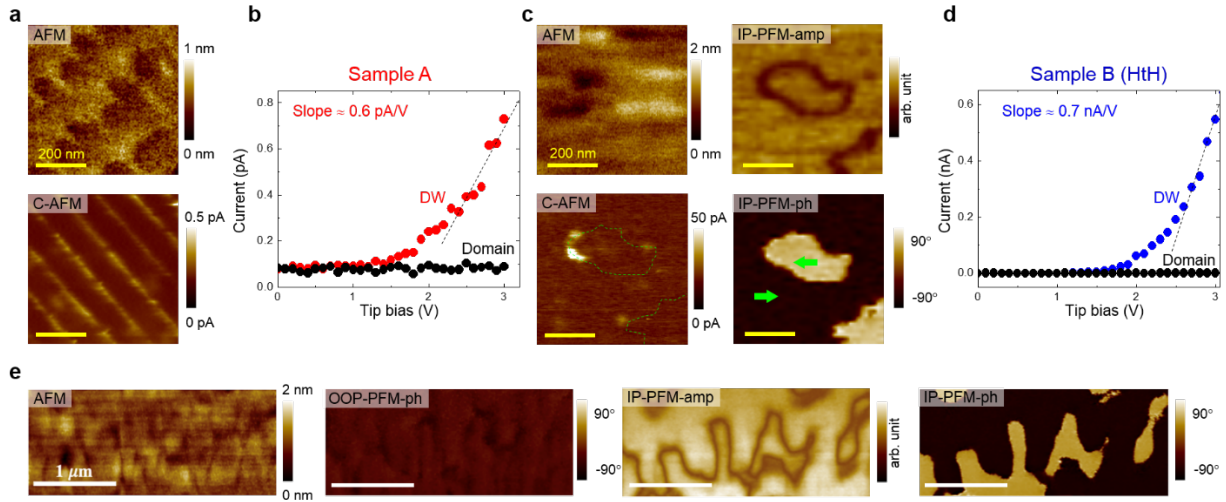

**Figure S3**. (**a**) AFM and C-AFM images in Sample A. (**b**) *I-V* curve on the DW and domain in Sample A. The slope is ~ 0.6 pA/V. (**c**) AFM, C-AFM, and phase/amplitude of in-plane PFM images in Sample B. Dashed lines in the C-AFM image show the contour of domains from the PFM data. All scale bars in (a) and (c) are 200 nm. (**d**) *I-V* curve on a head-to-head segment of DW and domain in Sample B. The slope is ~ 0.7 nA/V. (**e**) AFM, out-of-plane (OOP) and in-plane (IP) PFM images of Sample B in a large field of view.

The dc conductivity of DWs in the two BFO samples can be estimated from the C-AFM data (Figure S3a). Figure S3b shows the typical *I-V* curve of Sample A when the tip is on top of the DW, from which the slope of ~ 0.6 pA/V at high tip biases can be extracted. Assuming that the tip-DW contact area is 5 nm × 2 nm and the cross-sectional length of the DW is ~ 200 nm (150 nm film thickness with 45° tilt), one can show that the dc conductivity of inclined DWs in Sample A is on the order of $10^{-2}$ S/m, which is quoted in Figure 1d in the main text. In contrast, only segments of the charged DWs in Sample B exhibit a much larger current, as shown in Figure S3c. The combined C-AFM and PFM images indicate that these high-current regions correspond to the 'head-to-head' polarization configuration. Following the same steps, the dc conductivity of vertical DWs is ~ 3 S/m, as estimated from the *I-V* curve in Figure S3d. The results are in good agreement with reports in the literature [S3, S4].

We also include here the AFM and PFM images (Figure S3e) of sample B with a large field of view. The irregular patterns such as meander and closed domains are similar to that in our previous work [S5]. Note that the absence of phase contrast in the out-of-plane PFM image indicates that only one of the three quasi-cubic components of the BFO polarization changes direction across the wall, i.e., they are 71° DWs.

## S4. Detailed analysis of MIM and PFM images in Sample B.

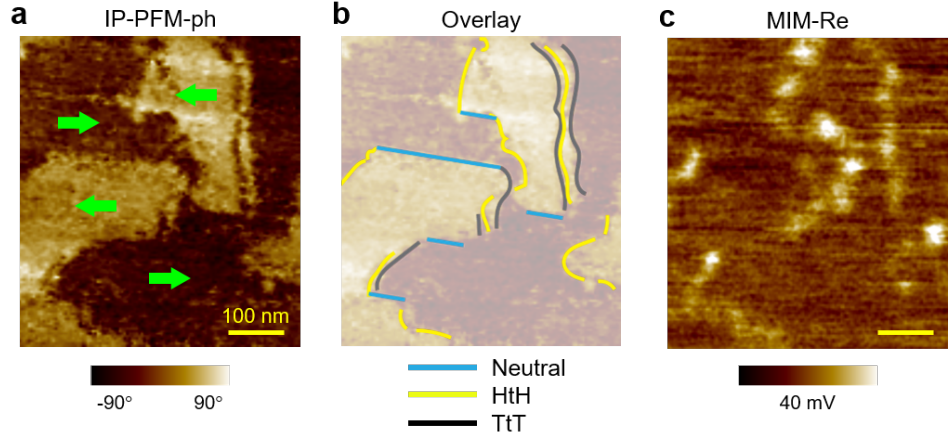

**Figure S4**. (**a**) In-plane PFM phase image in Sample B. The in-plane polarization directions of the domains are denoted by green arrows. (**b**) Shaded image of (a), with neutral, HtH, and TtT segments overlaid in the plot. (c) MIM-Re image of the same area. The scale bar is 100 nm.

A detailed analysis of the MIM-Re data in Sample B is shown in Figure S4. The domain structures are resolved by the in-pane PFM image in Figure S4a, from which the neutral, head-to-head (HtH), and tail-to-tail (TtT) segments of the DWs can be determined (Figure S4b). A comparison with the MIM-Re image in the same area (Figure S4c) shows that only the HtH sections with free-carrier accumulation display appreciable microwave response.

## S5. Cross-sectional TEM data of Sample B.

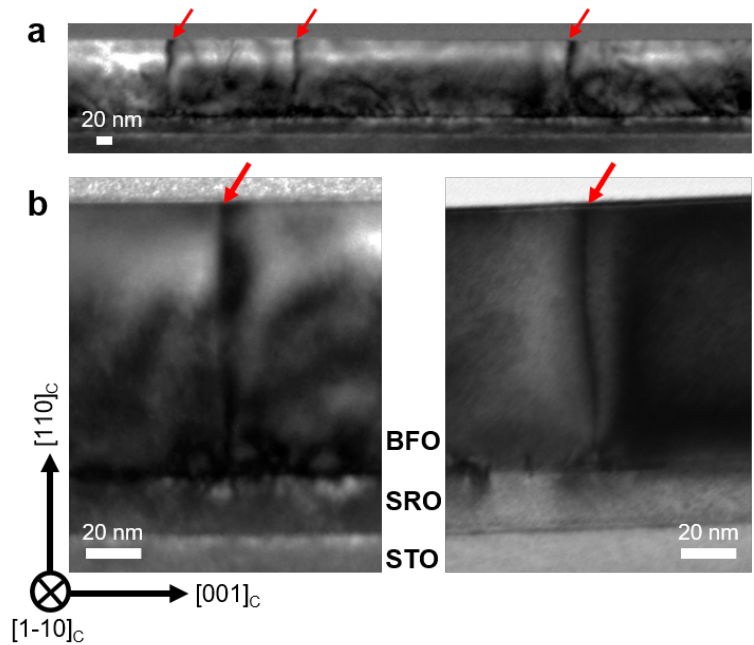

**Figure S5**. Dark-field transmission electron microscopy (TEM) images, low-magnification in (**a**) and high-magnification in (**b**), of $[110]_C$ BFO domain walls off the $[1\text{-}10]_C$ axis. It is clear that the charged walls are perpendicular to the surface, as illustrated in the schematic of Sample B.

An important assumption in the simulation and analysis of our results is that the head-to-head (HtH) DWs in our control sample are perpendicular to the surface. To experimentally verify this condition, we performed cross-sectional dark-field TEM imaging of Sample B. As shown in Figure S5, DWs in this sample are indeed vertical with respect to the film surface, consistent with our modeling in Figure 2 and Figure S7. On the other hand, it should be noted that the wall orientation may change following the contour of a closed domain in Sample B and the imaged section may not have the HtH configuration. To that end, we also simulate the DW orientation angle by phase-field modeling, as detailed below in Supporting Information S6.

## S6. DW orientation in Sample B.

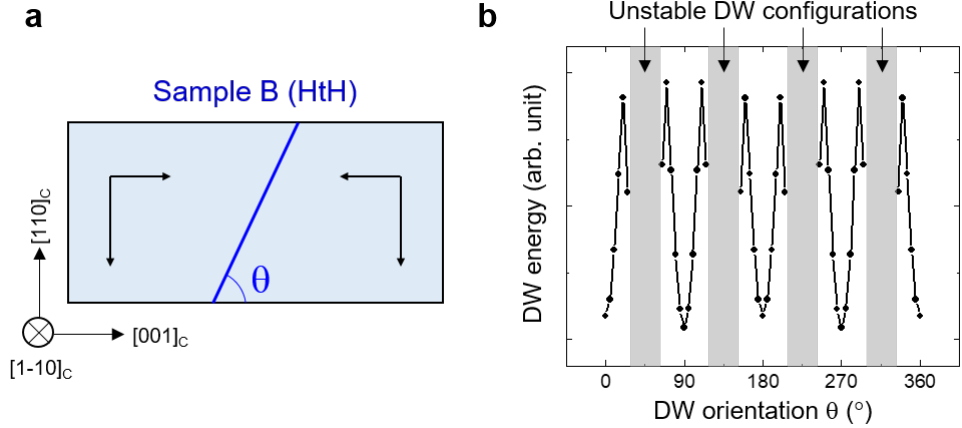

**Figure S6**. (**a**) Illustration of the domain wall orientation in the phase-field simulation. (**b**) Simulated DW energy as a function of the angle θ with respect to the $[001]_C$ axis, showing that the most stable configuration of the DW is at 90° (perpendicular to the film surface). The regions where DWs cannot be stabilized in the simulation are shaded in the plot.

Following the contour of a closed domain in Sample B, some parts of the wall may not orient vertically to the sample surface. In our experiment, however, the C-AFM and MIM images indicate that only the HtH sections of the DWs exhibit appreciable dc and ac conductivity, respectively. Consequently, we focus on the analysis of this particular configuration of the domain wall in Sample B. Using phase-field simulation, we have calculated the DW energy when the wall is rotated around the $[1\text{-}10]_C$ axis. As illustrated in Figure S6a, we denote the angle with respect to the $[001]_C$ axis as θ. The result in Figure S6b clearly indicates that θ = 90°, i.e., DW perpendicular to the sample surface, is the most favorable configuration in terms of the DW energy. Note that we have turned off the electrostatic interaction in the calculation due to the charged wall and only considered the elastic condition. In reality, it is possible that the wall may become rough in the microscopic length scale in order to accommodate the polarization charges. This minor effect, however, will not change our overall discussion of the HtH DW orientation in Sample B.

## S7. Stress and polarization fields in both samples.

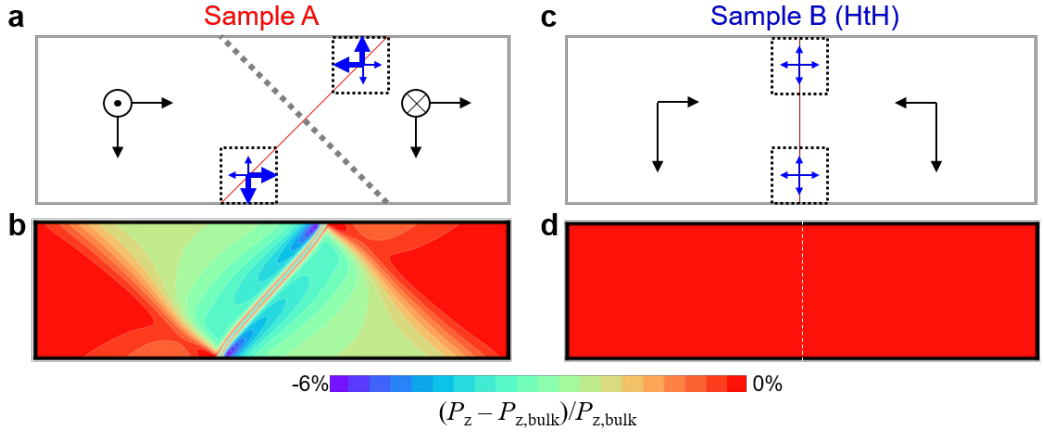

**Figure S7**. (**a**) Illustration of strain field near the inclined DW in Sample A. The dashed line indicates the mirror plane perpendicular to the wall. The polarization components are shown on both sides of the DW. Arrows in the dotted boxes represent the strength of stress along different directions. The net effect causes the DW to bend towards the surface normal. (**b**) Simulated out-of-plane polarization $P_z$ in Sample A, showing the imbalanced $P_z$ on two sides of the wall. (**c**) Illustration of strain field near the HtH section of the vertical DW in Sample B. The stress is balanced on two sides of the wall. (**d**) Simulated out-of-plane polarization in Sample B, showing the same $P_z$ as that in the bulk. The dashed line shows the unperturbed wall.

Figure S7 shows the results of our Ginzburg-Landau analysis on both samples [S6]. For Sample A, the asymmetric stress (dashed boxes in Figure S7a) near the film surface leads to deviation of the DW from its orientation in the bulk. The resultant asymmetric strain induces imbalanced $P_z$ on two sides of the wall via electrostriction, as seen in Figure S7b. For Sample B, however, the DW is perpendicular to the film surface, therefore no asymmetry in the stress/strain (Figure S7c) or polarization (Figure S7d) is induced around the wall.

## S8. Details of the phase-field modeling.

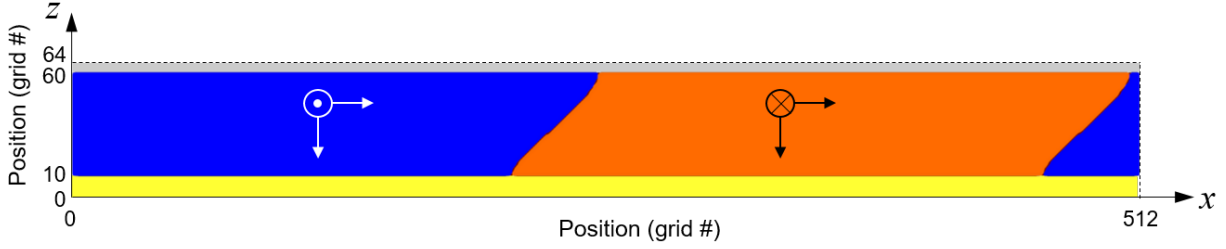

**Figure S8**. Configuration for the dynamical phase-field simulation of Sample A.

A phase-field model taking into account the polarization dynamics [S7] was used to simulate the domain and domain wall response in the BFO thin films under an ac electric field. The dynamic response of the local polarization field ***P(r)***, where ***r*** is the position vector, is described by a modified time-dependent Ginzburg-Landau equation with an additional term of second-order time-derivative of ***P*** accounting for its intrinsic oscillation, i.e.,

$$\mu \frac{\partial^2 \boldsymbol{P}}{\partial t^2} + \gamma \frac{\partial \boldsymbol{P}}{\partial t} + \frac{\delta F}{\delta \boldsymbol{P}} = 0 \tag{1}$$

where $\mu$ and $\gamma$ are kinetic coefficients related to the domain wall mobility. The equation was numerically solved using a semi-implicit Fourier spectral method. $F = F_{\text{landau}} + F_{\text{gradient}} + F_{\text{electric}} + F_{\text{elastic}}$ is the total free energy of the ferroelectric BFO. $F_{\text{landau}}$, $F_{\text{gradient}}$, and $F_{\text{electric}}$ are the ferroelectric landau free energy, ferroelectric gradient energy, and electrostatic energy, respectively [S8, S9]. The elastic energy $F_{\text{elastic}}$ is expressed as follows.

$$F_{\text{elastic}} = \int \frac{1}{2} c_{ijkl} (\varepsilon_{ij} - \varepsilon_{ij}^0)(\varepsilon_{kl} - \varepsilon_{kl}^0) d\boldsymbol{r}^3 \tag{2}$$

where **c** is the elastic stiffness tensor and $\varepsilon^0_{ij} = Q_{ijkl}P_k P_l$ is the stress-free strain related to the local ferroelectric order (**Q** denoting the electrostrictive coefficient tensor). The time step in the simulation is 0.01 ps. The polarization, permittivity, stiffness, and electrostrictive parameters of BFO were taken from Ref. [S10]. The kinetic coefficients $\mu = 10^{-16}$ J·m/A$^2$ and $\gamma = 10^{-4}$ J·m/(A$^2$·s) were used in the simulation. In particular, the effective damping coefficient $\gamma$ corresponds to a bulk dielectric loss of tan$\delta \sim 0.01$ at 1 GHz, which is consistent with the literature. Figure S8 shows the geometry of our 2D simulation, with 512 grids in the *x*-direction and 64 grids in the *z*-direction. Each grid here represents 0.4 nm. The BFO thin film, 50 grids in height, is terminated by 4 grids of vacuum on the top surface and 10 grids of conductive substrate on the bottom surface. The periodic boundary condition is applied in the $x$-direction. Two types of external potential were used in the simulation. For Figure 3b in the main text, a uniform ac electric field $E = E_0 \sin(2\pi t/T)$, where $E_0 = 10^5$ V/m and $T = 1$ ns, is applied between the top surface and the substrate. For Figure 3c in the main text, a Lorentz distribution of tip-like electric field (maximum field $10^5$ V/m) is scanned across the top surface. Finally, the displacement current density $\boldsymbol{j}_P = \partial \boldsymbol{P}/\partial t$ is simulated by the dynamical phase-field modeling and the time-averaged dielectric loss density $\partial \boldsymbol{P}/\partial t \cdot \boldsymbol{E}$ is plotted in Figure 3d in the main text.

**References:**

[S1]   X. Wu, U. Petralanda, L. Zheng, Y. Ren, R. Hu, S.-W. Cheong, S. Artyukhin, and K. Lai, *Sci. Adv.* **2017**, 3, e1602371.

[S2]   K. Lai, W. Kundhikanjana, M. Kelly, and Z. X. Shen, *Rev. Sci. Instrum.* **2008**, 79, 063703.

[S3]   J. Seidel, L. W. Martin, Q. He, Q. Zhan, Y. H. Chu, A. Rother, M. E. Hawkridge, P. Maksymovych, P. Yu, M. Gajek, N. Balke, S. V. Kalinin, S. Gemming, F. Wang, G. Catalan, J. F. Scott, N. A. Spaldin, J. Orenstein, and R. Ramesh, *Nat. Mater.* **2009**, 8, 229.

[S4]   A. Crassous, T. Sluka, A. K. Tagantsev, and N. Setter, *Nat. Nanotech.* **2015**, 10, 614.

[S5]   Y.-H. Chu, M. P. Cruz, C.-H. Yang, L. W. Martin, P.-L. Yang, J.-X. Zhang, K. Lee, P. Yu, L.-Q. Chen, and R. Ramesh, *Adv. Mater.* **2007**, 19, 2662.

[S6]   P. Chen, L. Ponet, K. Lai, and S. Artyukhin, arXiv:1907.12989.

[S7]   H. Akamatsu, Y. Yuan, V. A. Stoica, G. Stone, T. Yang, Z. Hong, S. Lei, Y. Zhu, R. C. Haislmaier, J. W. Freeland, L.-Q. Chen, H. Wen, and V. Gopalan, *Phys. Rev. Lett.* **2018**, 120, 096101.

[S8]   Y. L. Li, S. Y. Hu, Z. K. Liu, and L. Q. Chen, *Appl. Phys. Lett.* **2001**, 78, 3878.

[S9]   Y. L. Li, S. Y. Hu, Z. K. Liu, and L. Q. Chen, *Appl. Phys. Lett.* **2002**, 81, 427.

[S10]  R. K. Vasudevan, A. N. Morozovska, E. A. Eliseev, J. Britson, J.-C. Yang, Y.-H. Chu, P. Maksymovych, L. Q. Chen, V. Nagarajan, and S. V. Kalinin, *Nano Lett.* **2012**, 12, 5524.